\documentclass[%
reprint,
showpacs,
 amsmath,amssymb,
 aps,
]{revtex4-1}

\usepackage{graphicx}% Include figure files
\usepackage{dcolumn}% Align table columns on decimal point
\usepackage{bm}% bold math

\newcommand{\PP}{\mathbf{P}}
\newcommand{\EE}{\mathbf{E}}

\newcommand{\la}{\langle}
\newcommand{\ra}{\rangle}

\begin{document}

%\preprint{APS/123-QED}

\title{Does good memory help you win games?}

\author{James Burridge}
\affiliation{Department of Mathematics, University of Portsmouth,
  Portsmouth, PO1 2UP, UK}

\author{Yu Gao}
\author{Yong Mao}
\affiliation{School of Physics and Astronomy, University of
  Nottingham, Nottingham, NG7 2RD, UK}

\date{\today}

\begin{abstract}
We present a simple game model where agents with different memory lengths compete for finite resources. We show by simulation and analytically that an instability exists at a critical memory length, and as a result, different memory lengths can compete and co-exist in a dynamical equilibrium. Our analytical formulation makes a connection to statistical urn models, and we show that temperature is mirrored by the agent's memory. Our analysis is easily generalisable to many other game models with implications that we briefly discuss.
\end{abstract}


\maketitle

\emph{Introduction} -- All successful forms of life must eventually
engage in competition for resources.  The equilibrium analysis of
these competitions began with von Neumann \cite{Neu53} and Nash
\cite{Nash51}. The theory of games has since found applications in
genetics, ecology, economics and sociology
\cite{May82,Hof1,Mye91,Far09}. Computational implementation of games
leads to agent-based models, which may be of particular importance in
understanding the behaviour of financial systems
\cite{Far09,Chak15}. For example, the particularly successful minority game
model \cite{Cha97,Cha05,Cav99,Gar00}, captures the competition between
intelligent agents with a restricted form of memory.
Recent work suggest such games may be generalised leading to clearly
separated regimes of behaviour \cite{Gal13}. In general, understanding the complex collective behaviour arising from the non-linear interactions between individuals is a major challenge for statistical physics \cite{Bou02,Man99}.

In this Letter we present a simple game model: in each round,
individual agents pick one of two urns, each offering a stochastic
yield to be shared by the pickers. An agent has a memory of these
payouts for the previous $\tau$ rounds to aid its decision.
Such stochastic yield sharing arises in animal foraging behaviour and stock trading \cite{Saa13}, where the urns may represent different prey species, foraging patches, or stocks. Some form of intelligence is essential in order to compete \cite{Bel97}. As with the minority game \cite{Cha97}, agents' memory in our model is a tool for decision making. However, unlike the minority game, memory in our model is used to make direct estimates of the highest paying choice, rather than to second guess opponents' next moves.
%Mapping to the minority game would require many urns each representing a %different response to a fictitious history \cite{Cav98}.
In common with the thermal minority game \cite{Cav99}, dynamical urn models \cite{God01}, and some evolutionary games \cite{Tra07}, our agents have a ``temperature'', which captures the level of noise in the switches they make in search of yields. We find that the additional noise inherent in their finite memory samples, which is greater for shorter memories, leads the system to behave as if its agents have a higher temperature. Therefore, increasing memory ``cools'' the system. However, at a critical memory, a Hopf bifurcation \cite{Ern09} emerges producing stable cycles in the numbers of agents in each urn. Perhaps not surprisingly, a long memory is advantageous, but the presence of these cycles allows short memory agents to compete, and a mixed memory system will evolve toward the bifurcation point.  Our theoretical formulation follows that of statistical urn models such as Ehrenfest's dog flea model \cite{Ehr07}, which
played an important role in the early development of statistical mechanics, and more recently allowed analytical investigation of effects such as slow relaxation and condensation in nonequilibrium statistical mechanics \cite{God01}.

Whilst we have restricted this Letter to a yield sharing game, the analysis may be carried through for any social system where agents switch between strategies using their memory to determine the optimal choice. Simple examples include the Hawk Dove \cite{May82} and Rock Scissors Paper \cite{Wan14} games. Our formulation could also accommodate the inclusion of topological effects and agent interactions \cite{Cac08}. However, even without such complexities, urn models exhibit a remarkable range of non-equilibrium behavior that is connected with temperature \cite{God01}. Our analysis suggests that effects such as condensation and the emergence of order \cite{Cas09}, which have social interpretation, may also have a connection with memory \cite{Cac08}, but that long memory can also
introduce instability.

\emph{Model definition} -- Consider the case of two urns and a total of $n$ agents. We let the urn yields, $U_1(t), U_2(t)$, at round $t$ be random variables uniformly distributed on $[0,\omega n]$ and $[0,n]$ respectively, where $\omega>1$ so that urn 1 yields more on average than urn 2. We allow agents access to
the arithmetic mean of the last $\tau$ payoffs, but we note that other forms of sampling could be used.  Letting $\phi_t$ be the fraction of agents in urn $1$, then the difference in the average payoffs between urn 1 and urn 2 is
\begin{equation}
\label{Delta}
\Delta_t := \frac{1}{\tau}\sum_{s=0}^{\tau-1} \left[\frac{U_2(t-s)}{n
    (1-\phi_{t-s})} - \frac{U_1(t-s)}{n \phi_{t-s}}\right].
\end{equation}
We refer to $\tau$ as the ``memory length'' of the agents. Agent
dynamics is encoded in transition probabilities between urns, which
are deterministic functions of $\Delta_t$. At each round, each agent
will switch urns using the probabilities
\begin{align}
\label{W12}
W_{1 \rightarrow 2}(\Delta) &=  \frac{\epsilon}{2}\left[1+\tanh(\beta \Delta)\right] \\
W_{2 \rightarrow 1}(\Delta) &= \frac{\epsilon}{2}\left[1-\tanh(\beta \Delta)\right].
\label{W21}
\end{align}
The parameter $\beta$, the ``inverse temperature'', captures the degree of stochasticity with which agents make choices. For finite $\beta$, agents may decide to switch strategies even though their estimate of the payoff difference is unfavourable. In the limit $\beta \rightarrow \infty$ agents will only move if their estimate of the payoff difference indicates that the move is favourable. The parameter $\epsilon$ controls the rate at which strategy switching takes place
compared to the rate at which yield information arrives. It may also be seen as the frequency with which opportunities to switch strategy arise. In the limit
$\epsilon \rightarrow 0$, at most one agent will move at each round.

\begin{figure}
\begin{center}
\includegraphics[width=8.5 cm]{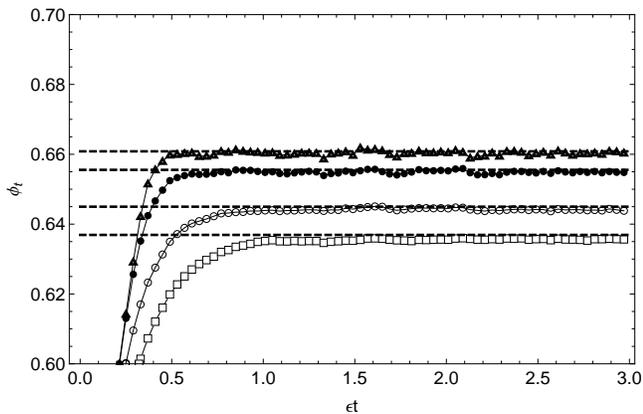}
\caption{ Evolution of $\phi_t$ when $n=10^6$, $\omega = 2$, $\beta=10$
  and $\epsilon=10^{-6}$. Memory values are $\tau \in \{5,10,50,500\}$ (squares, circles, dots, triangles). Dashed lines are analytical equilibrium values (see equation (\ref{phiEq})). \label{equilibs}}
\end{center}
\end{figure}

\begin{figure}
\begin{center}
\includegraphics[width=8.5 cm]{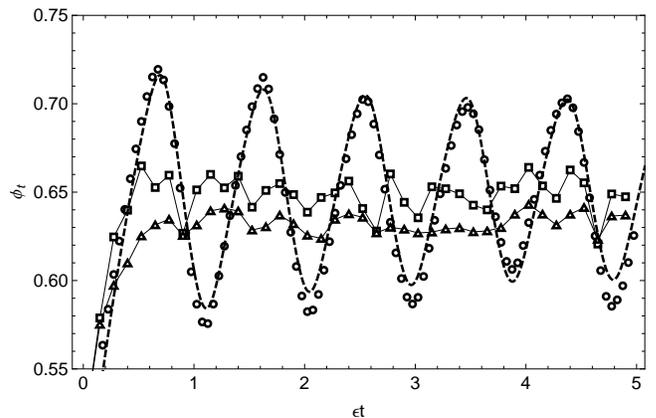}
\caption{ Evolution of $\phi_t$ when $n=10^6$, $\omega = 2$, $\beta=5$
  and $\epsilon=10^{-3}$. Memory values are $\tau \in \{5,50,500\}$ (triangles, squares, circles). Dashed line is solution to equation (\ref{delayEqn}) when $\tau=500$ and $\omega, \beta, \epsilon$ are as above.\label{phiOsc}}
\end{center}
\end{figure}

\emph{Simulation (instability)} --
We simulate the model for a series of values of $\tau$ when $n=10^6$. Two different values of $\epsilon$ are used; in Figure \ref{equilibs} we have $\epsilon^{-1} = 10^6 \gg \tau$ and in Figure \ref{phiOsc} we have $\epsilon=10^{-3}$. For $\epsilon = 10^{-6}$, the expected number of moves at
each step is $< 1$, and $\phi$ appears very stable. For larger $\epsilon$, $\phi$ experiences much larger fluctuations about the steady state value, driven by the yield process. For shorter memory values these fluctuations are random, but as $\tau$ approaches $\epsilon^{-1}$, periodic oscillations appear and dominate. The appearance of these stable oscillations at critical memory, $\tau_c$, is known as a Hopf bifurcation \cite{Ern09}. By allowing agents access only to the mean of their memory, we implicitly assume that changes in the expected payoff over the course of their memory, brought about by oscillations, are too subtle for them to infer from noise.

\begin{figure}
\begin{center}
\includegraphics[width=8.5 cm]{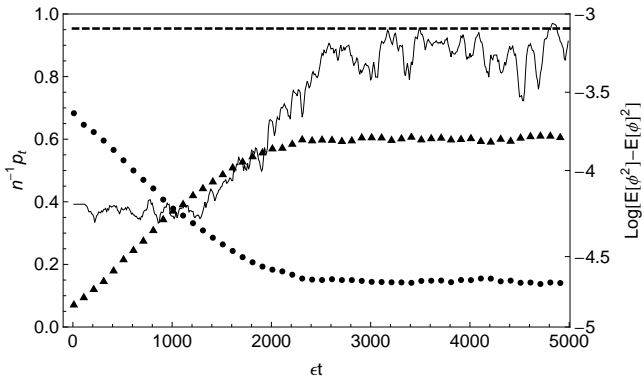}
\caption{ Scaled populations $p(t) := p^\tau_1(t) + p^\tau_2(t)$ for
$\tau=10$ (circles) and $\tau=1000$ (triangles) when $n=p(0)=10^6$, $\epsilon= 10^{-3}$, $\beta =4$ with $ \gamma  = 10^{-4}$ and $\delta = 2 \times 10^{-4}$.  Also shown (thin black line) is evolution of variance of $\phi_t$ during population dynamics simulation. Straight dashed line shows variance of homogeneous population with the same $\epsilon, \beta, \omega$  values at critical memory $\tau_c$. Note: rapid initial equilibration of population values (bringing birth and death into balance) is not visible on time scale of plot. \label{popDyn}}
\end{center}
\end{figure}

\emph{Simulation (coexistence)} --
We now investigate how agents with two different memories compete against one another by interpreting the payoff as reproduction rate. We define $\delta$ and $\gamma$ as the rates of death, and reproduction per unit payoff, respectively. Reproduction is assumed to occur before death in each round, but in practice the probability of any one agent reproducing and dying in the same round is extremely small for the $\gamma, \delta$ values we choose. Letting $p^\tau_i(t)$ be the number of agents with memory $\tau$ in urn $i$ at time $t$ we set the probability of birth for an agent in urn $i$ to be
\begin{equation}
\PP(\text{birth}) = \frac{\gamma U_i(t)}{\sum_\tau p_i^\tau(t)}.
\end{equation}
The death probability for each agent is set equal to $\delta$. If populations are fixed in size and the system is not in an oscillatory state, then we expect that in equilibrium the longer memory agents will dominate the high yielding urn. Their long memory allows them to perceive smaller statistical advantages that are obscured by noise for the short memory agents. Using the thermodynamic analogy, the higher temperature (shorter memory) agents are more likely to make moves
which leave them in an urn with a lower expected payoff, corresponding to a higher ``energy'' state.  Above zero temperature, and in the absence of oscillations, the high yield urn will be under-exploited, placing high memory agents at an advantage. This effect can be observed in Figure \ref{popDyn} where we have simulated a mixed population of two memories $\tau \in\{10,1000\}$ beginning with a ratio of 10:1 short memory to long memory agents. We see that initially the advantage afforded the long memory agents causes their population to grow, whereas the short memory agents reduce in number. Were this advantage to be sustained indefinitely then we would expect the short memory agents to eventually disappear, but in fact the populations stabilize. This effect appears because the long memory agents cause oscillations to develop once they are in sufficiently high concentration. In the presence of oscillations the short memory agents have an advantage because they can quickly observe opportunities offered by the oscillating payoffs. We therefore expect the system to evolve to the point where oscillations are just beginning to form. We may observe this evolution by making use of the variance of $\phi_t$ as an order parameter which captures proximity to the Hopf bifurcation point. In Figure \ref{popDyn} we see that at a critical ratio of short to long memory agents, the variance climbs rapidly, stabilizing just
below the value seen in a system where all agents have memory $\tau_c$ but all other parameters are equal. In this way the Hopf bifurcation may be viewed as a self organized state.

\emph{Analysis (equilibrium)} -- We consider the behaviour of the model as $\epsilon \rightarrow 0$, allowing us to view it as an urn model in the Ehrenfest class \cite{God01} where agents independently make transitions using state ($\phi_t$) dependent probabilities. Provided $\tau \ll \epsilon^{-1}$, the fraction $\phi_t$ may be approximated by a constant $\phi$ during the window over which payoff averaging takes place. In this case, by the central limit
theorem, the marginal distributions of $\Delta_t$ for each $t$ are
approximately normal $ N(\bar{\Delta}, \sigma^2/\tau)$ where, from
(\ref{Delta})
\begin{align}
\label{DeltaBar}
\bar{\Delta}(\phi,\omega) &:= \frac{1}{2} \left( \frac{1}{1-\phi} - \frac{\omega}{\phi} \right) \\
\frac{\sigma^2(\phi,\omega)}{\tau} &:=  \frac{1}{12 \tau} \left[ \frac{\omega^2}{\phi^2} + \frac{1}{(1-\phi)^2} \right].
\label{sigmaSq}
\end{align}
We now introduce a intermediate time scale $T$ satisfying $\tau \ll T
\ll \epsilon^{-1}$ and define the time average $\la \cdot \ra$, over a
window of length $T$
\begin{equation}
\la W_{i \rightarrow j}(\Delta) \ra (t) := \frac{1}{T} \sum_{s = t-T+1}^t W_{i \rightarrow j}(\Delta_s).
\end{equation}
This average is a random variable which, for constant $\phi$, has expected value $ \EE[ W_{i \rightarrow j}(\Delta) ]$ where the expectation is taken over the marginal distribution of $\Delta$. The condition $\tau \ll T \ll \epsilon^{-1}$ ensures that $\phi$ is approximately constant over the window and that the variance of $\la W_{i \rightarrow j}(\Delta) \ra $ is proportional to $T^{-1}$ (because $\Delta_{t_1}$ and $\Delta_{t_2}$ are dependent only when $|t_2-t_1| < \tau \ll T$). As $\epsilon \rightarrow 0$, then assuming $T$ is sufficiently large, the probability that an agent will make a transition $i \rightarrow j$ during interval $T$ approaches $T \la W_{i \rightarrow j}(\Delta) \ra \approx T \EE[ W_{i \rightarrow j}(\Delta) ]$, equivalent to a memoryless (Ehrenfest class) model where transition probabilities (\ref{W12},\ref{W21}) are replaced with their expectations $\EE[ W_{i \rightarrow j}(\Delta) ] $. Averaging over the normally distributed difference $\Delta$ we find that
\begin{equation}
\label{WB}
\la W_{1 \rightarrow 2}(\Delta) \ra \approx \EE[ W_{1 \rightarrow
    2}(\Delta) ] \approx \frac{\epsilon}{2} \left[1+ \tanh(\alpha
  \bar{\Delta}) \right]
\end{equation}
where
\begin{equation}
\alpha = \sqrt{\frac{2 \tau \beta^2}{2 \tau + \pi \beta^2 \sigma^2}}.
\end{equation}
To obtain this result, we have made the approximation $\tanh(\beta \Delta)
\approx \text{erf}(\sqrt{\pi} \beta \Delta/2) $, allowing us to make use of the exact relationship $\EE[\text{erf}(\sqrt{\pi}\beta \Delta/2)]= \text{erf}(\sqrt{\pi} \alpha \bar{\Delta}/2)$. The constant $\alpha$ acts as an effective inverse temperature and we see that increasing $\tau$ ``cools'' the system closer to the inverse temperature $\beta$, and in the limit $\beta
\rightarrow \infty$, $\alpha \propto \sqrt{\tau}$. To complete our
analogy to a thermal urn model we now write the probability of finding
the agents in a particular arrangement, or microstate, $i$, such that
a fraction $\phi$ are in urn 1, as $p_i(\phi) \propto e^{-\alpha E}$
where $E$ is an ``energy'' function. Considering two microstates separated by a single transition, and defining $\delta \phi = 1/n$, then detailed balance requires that in equilibrium $2 \alpha\bar{\Delta} = \partial_\phi (\alpha E) \delta \phi$. This condition allows $E(\phi)$ to be computed, in principle, by integration. A closed form approximation $E(\phi) \approx -n  \ln \left[\phi^\omega(1-\phi)\right]$ is obtained by noting that $\alpha$ depends weakly on $\phi$ compared to $E$ so that $\partial_\phi (\alpha E) \approx \alpha \partial_\phi E$. Summing over all microstates corresponding to macrostate $\phi$ we have a Boltzmann probability distribution for $\phi$
\begin{equation}
\PP(\phi) = \frac{n!}{(n\phi)!(n(1-\phi))!} \frac{e^{-\alpha(\phi) E(\phi)}}{\mathcal{Z}},
\end{equation}
where $\mathcal{Z}$ is the partition function. Taking the
thermodynamic limit $n \rightarrow \infty$, and making use of
Stirling's approximation, we find that the most likely (maximum entropy)
fraction, $\bar{\phi}$, satisfies:
\begin{equation}
\label{equilib}
\frac{1}{2n} \frac{\partial}{\partial \phi} \ln \PP(\phi)= \alpha \bar{\Delta}  - 2\phi + 1=0.
\end{equation}
As the memory increases and the system cools we expect the agents to
arrange themselves so that yields are shared more fairly. We therefore
linearize (\ref{equilib}) about the perfectly fair state, $\phi =
\omega/(1+\omega)$,  where agents in both urns receive the same
expected payoff, finding that
\begin{equation}
\bar{\phi} \approx \frac{ f(\tau)+  \frac{\beta (1+\omega)^2}{2}}{2 f(\tau) + \frac{\beta (1+\omega)^3}{2 \omega}},
\label{phiEq}
\end{equation}
where $ f(\tau) = \sqrt{1+\pi \beta^2 (1+\omega)^2/(12 \tau)}$. The accuracy of this approximation is verified in Figure \ref{equilibs}. For larger values of $\epsilon$ (Figure \ref{phiOsc}) agents move more quickly so the averaging effect (\ref{WB}) damps fluctuations in transition rates less strongly, creating larger fluctuations in $\phi_t$.  For finite $\beta$ the system cannot reach perfect fairness for any memory length, but in the limit $\beta \rightarrow \infty$ where the transition probabilities (\ref{W12},\ref{W21}) become step functions,
we have that:
\begin{equation}
\bar{\phi} \approx \frac{\omega}{\omega +1}  \left[1-\frac{\sqrt{\pi} (\omega
   -1)}{\sqrt{3 \tau } (\omega +1)^2} + \mathcal{O}(\tau^{-1})\right].
\end{equation}
From this we see that the distance away from the fair state decreases
as $\tau^{-1/2}$ as the memory of the agents becomes large. However,
we now show why increasing $\tau$ too far, when $\epsilon$ is finite, destabilizes the system.

\emph{Analysis (instability)} -- As $\tau$ increases,
fluctuations in $\Delta_t$ due to the yield process are reduced but for finite $\epsilon$ we can no longer treat $\phi_t$ as a constant over the averaging
window. It is instructive, therefore, to study the effect of
variations in $\phi_t$, neglecting the variations in yield. Promoting
$t$ to a continuous variable and replacing the urn yields with their
mean values we have
\begin{equation}
\Delta_t \approx \frac{1}{2\tau} \int_{t-\tau}^t \left[ \frac{1}{1-\phi_{s}} - \frac{\omega}{\phi_{s}} \right] ds.
\end{equation}
We then approximate the evolution of $\phi_t$ using the following delay differential equation:
\begin{equation}
\dot{\phi}_t =  (1-\phi_{t}) W_{2 \rightarrow 1}(\Delta_t) - \phi_t W_{1 \rightarrow 2}(\Delta_t).
\label{delayEqn}
\end{equation}
A numerical solution to this equation is shown in Figure
\ref{phiOsc}, along with simulation results using the same parameter
values. The oscillations in the simulation are accurately captured by (\ref{delayEqn}) but the stochastic yield disrupts their perfect periodicity. To discover the parameter values at which stable oscillations develop we linearize equation (\ref{delayEqn}) by writing $\phi_t = \bar{\phi} + \psi_t$ where $\psi_t$ are small fluctuations and $\bar{\phi}$ is the
constant fixed point, not necessarily stable, of equation (\ref{delayEqn}). In terms of these new variables
\begin{equation}
\Delta_t \approx \bar{\Delta}(\bar{\phi},\omega) +
6\frac{ \sigma^2(\bar{\phi},\sqrt{\omega})}{\tau} \int_{t-\tau}^t \psi_s ds
\end{equation}
where the functions $\bar{\Delta}$ and $\sigma^2$ are defined in
(\ref{DeltaBar}) and (\ref{sigmaSq}). After expanding the $\tanh$
functions in the transition rates to first order about
$\bar{\Delta}(\bar{\phi},\omega)$, we obtain the following linear
delay equation
\begin{equation}
\dot{\psi}_t = -\epsilon \left[ \psi_t + \frac{A}{\tau} \int_{t-\tau}^t \psi_s ds \right]
\label{lin}
\end{equation}
where $A = 3 \beta \text{sech}^2[\beta \bar{\Delta}(\bar{\phi},\omega)]
\sigma^2(\bar{\phi},\sqrt{\omega})$. To determine the stability of
this equation we introduce an exponential trial solution $\psi_t =
e^{\lambda t}$ where $\lambda = x + i y$. Substitution into equation
(\ref{lin}) yields a characteristic equation with real and imaginary
parts given by
\begin{align}
\label{re}
x^2 - y^2 + \epsilon x + \frac{\epsilon A}{\tau} \left( 1- e^{-\tau x} \cos \tau y \right) &= 0\\
2xy +\epsilon y  + \frac{\epsilon A}{\tau} e^{-\tau x} \sin \tau y &= 0.
\label{im}
\end{align}
For sufficiently small memory, $\tau$, the real part, $x$, of the
solutions to $(\ref{re})$ and $(\ref{im})$ is negative so the fixed
point $\bar{\phi}$ is stable. As we increase $\tau$, $\lambda$ crosses
through the imaginary axis, creating a switch to instability with
oscillations of exponentially increasing magnitude. Although the full
equation (\ref{delayEqn}) shares this transition to instability, we
find that the resulting oscillations are bounded. The appearance of
these stable oscillations as $\tau$ passes through a critical value,
which we denote $\tau_c$, constitutes the Hopf Bifurcation
\cite{Ern09}. To compute $\tau_c$ we set $x=0$ in equation (\ref{im})
so that $\text{sinc}(\tau y) = A^{-1}$. Expanding the sinc function to
second order about its root at $\pi/\tau$ and solving the resulting
quadratic we find that
\begin{equation}
y \approx \frac{\pi}{2 \tau} \left( 3 - \sqrt{1-4 A^{-1}}\right) := \frac{\kappa}{\tau}
\end{equation}
which defines a new constant $\kappa$. Substitution of this solution
into (\ref{re}), yields the following expression for the critical
memory length
\begin{equation}
\tau_c = \frac{\kappa^2}{\epsilon A (1-\cos \kappa)}.
\label{tauc}
\end{equation}
For example, for the parameter values used in Figure \ref{phiOsc}, we have $\tau_c \approx 400$, whereas the relevant critical value for Figure \ref{equilibs} is $\tau_c = 1.8 \times 10^{5}$. These values are in excellent agreement with simulations.

\emph{Conclusion}-- We have introduced a simple thermal urn model
of competition between agents with memory. Increasing memory allows
agents to more accurately determine the most productive strategy, and
reduces the temperature of the model. However, if a sufficiently high
concentration of long memory agents is present a limit cycle appears
which reduces the competitiveness of long memory agents, leading to
self organized Hopf bifurcation in a mixed memory model. The
simplicity of our model, its connection to classical urn models,
together with the fact that limit cycles arise naturally suggest it
might be fruitfully generalized, and employed to study a variety of
different games. For example our approach may be applied to the Rock
Scissors Paper game \cite{May82}, where agents, interacting pairwise,
recall their last $\tau$ interactions. Although a larger memory
provides better statistical data on the optimal strategy, at critical
memory a limit cycle emerges about Nash equilibrium, destroying this
competitive advantage \cite{Burr15}. Other natural extensions include the introduction of multiple urns to represent, for example, different financial stocks. In this case we would expect more complex patterns of oscillation \cite{Man04}. Experimental research into the nature of human and animal memory \cite{Est72,Kil01,Coo06,Ave11} places emphasis on the ``forgetting function'' which describes how memories decay with time. Such a function, or greater powers of statistical inference, could be naturally incorporated into our analysis, and its effects on stability explored.


\begin{thebibliography}{9}

\bibitem{Neu53}
J. ~Von Neumann and O. ~Morgenstern, \textit{Theory of Games and Economic Behaviour}, Princeton University Press (1953).

\bibitem{Nash51}
J. ~Nash, Annals of Mathematics, \textbf{54}, 286-295 (1951).

\bibitem{May82}
J. ~Maynard Smith, \textit{Evolution and the theory of games}, Cambridge University Press (1982).

\bibitem{Mye91}
R.~B.~Myerson, \textit{Game Theory: Analysis of Conflict}, Harvard University Press (1991).

\bibitem{Hof1}
J.~Hofbauer and K.~Sigmund, \textit{Evolutionary Games and Population Dynamics}, Cambridge University Press (1998).

\bibitem{Far09}
J.~D.~Farmer and D.~Foley, Nature, \textbf{460}, 685-686 (2009).

\bibitem{Chak15}
A. ~Chakraborti, D. ~Challet, A. ~Chatterjee, M. ~Marsili, Y.-C. ~Zhang and B. K. ~Chakrabarti, Physics Reports, \textbf{552} 1-25 (2015).

\bibitem{Cha97}
D.~Challet and Y.-C.~Zhang, Physica A, \textbf{246}, 407-418 (1997).

\bibitem{Cha05}
D.~Challet, M.~Marsili and Y.~C.~Zhang, \textit{Minority Games}, Oxford University Press (2005).

\bibitem{Cav99}
A. ~Cavagna, J. P. ~Garrahan and I. ~Giardina and D. ~Sherrington, Phys. Rev. Lett. \textbf{83} 4429 (1999).

\bibitem{Gar00}
J. P. ~Garrahan, E. ~Moro and D. ~Sherrington, Phys. Rev. E. \textbf{62} R9 (2000).

\bibitem{Gal13}
T.~Galla and J.~D.~Farmer, PNAS, \textbf{110}, 1232 (2013).


\bibitem{Bou02}
J.-P.~Bouchaud, Physica A, \textbf{313}, 238-251 (2002).


\bibitem{Man99}
R.~Mantegna and H.~E.~Stanley, \textit{An Introduction to Econophysics}, Cambridge University Press (1999).


\bibitem{Saa13}
S. ~Saavedra, R. D. ~Mailmgren, N. ~Switanek and B. ~Uzzi, P. R. Soc. B. \textbf{280} 20122901 (2013).

\bibitem{Bel97}
C. ~Belisle and J. Cresswell, Theor. Popul. Biol. \textbf{52} 78-90 (1997).


\bibitem{Cav98}
A. ~Cavagna, Phys. Rev. E., \textbf{59} 3783 (1999).


\bibitem{God01}
C.~Godreche and J.~M.~Luck, J. Phys. Cond. Matt. \textbf{14} 1601-1615 (2002).

\bibitem{Tra07}
A. ~Traulsen, J. M. Pacheco and M. A. Nowak, J. Theor. Biol. \textbf{246} 522-529 (2007).

\bibitem{Ern09}
T. ~Erneux, \emph{Applied Delay Differential Equations}, Springer (2009).

\bibitem{Ehr07}
P.~Ehrenfest, Phys. Zeit., \textbf{8}, 311 (1907).

\bibitem{Wan14}
Z. ~Wang, B. ~Xu and H.J. ~Zhou, Scientific Reports \textbf{4} 5830 (2014).

\bibitem{Cac08}{
F. Caccioli, S. Franz and M. Marsili., J. Stat. Mech: Theory and Experiment, \textbf{7}, P07006 (2008)}

\bibitem{Cas09}
C. ~Castellano, S. ~Fortunato and V. ~Loreto, Rev. Mod. Phys. \textbf{81} (2009).

\bibitem{Mah08}
H. ~Mahmoud, \textit{Polya urn Modles}, CRC Press, Boca Raton (2008).

\bibitem{Man04}
P. ~Manfredi and L. ~Fanti, Economic Modelling, \textbf{21} 573-594 (2004).

\bibitem{Burr15}
J. ~Burridge, In Preparation.

\bibitem{Est72}
W. K. ~Estes, J. Am. Stat. Assoc, \textbf{67} 81-102 (1972).

\bibitem{Kil01}
P. ~Killeen, Psychon. B. Rev. \textbf{8} 18-43 (2001).

\bibitem{Coo06}
R. G. ~Cook and A. P. ~Blaisdell, Behav. Proc. \textbf{72} 255-264 (2006).

\bibitem{Ave11}
L. ~Averell and A. ~Heathcote, J. Math. Psychol. \textbf{55} 25-35 (2011).





\end{thebibliography}
\end{document}